\RequirePackage{amsthm}

\documentclass[pdflatex, sn-basic, Numbered]{sn-jnl}

\usepackage{graphicx}%
\usepackage{multirow}%
\usepackage{amsmath,amssymb,amsfonts}%
\usepackage{amsthm}%
\usepackage{mathrsfs}%
\usepackage[title]{appendix}%
\usepackage[table, x11names]{xcolor}
\usepackage{siunitx}
\DeclareSIUnit\px{px}
\usepackage{hyperref}
\usepackage{layouts}
\usepackage{textcomp}%
\usepackage{manyfoot}%
\usepackage{booktabs}%
\usepackage{algorithm}%
\usepackage{algorithmicx}%
\usepackage{algpseudocode}%
\usepackage{listings}%

\theoremstyle{thmstyleone}%
\theoremstyle{thmstyletwo}%

\theoremstyle{thmstylethree}%

\begin{document}

\title{Improving 3D deep learning segmentation with biophysically motivated cell synthesis}


\author*[1]{\fnm{Roman} \sur{Bruch}}\email{roman.bruch@kit.edu}
\author[2,3]{\fnm{Mario} \sur{Vitacolonna}}\email{m.vitacolonna@hs-mannheim.de}
\author[2,3]{\fnm{Elina} \sur{Nürnberg}}\email{e.nuernberg@hs-mannheim.de}
\author[2,4]{\fnm{Simeon} \sur{Sauer}}\email{s.sauer@hs-mannheim.de}
\author[2,3]{\fnm{Rüdiger} \sur{Rudolf}}\email{r.rudolf@hs-mannheim.de}
\author[1]{\fnm{Markus} \sur{Reischl}}\email{markus.reischl@kit.edu}

\affil*[1]{\orgdiv{Institute for Automation and Applied Informatics}, \orgname{Karlsruhe Institute of Technology}, \orgaddress{\street{Hermann-von-Helmholtz-Platz 1}, \city{Eggenstein-Leopoldshafen}, \postcode{76344}, \country{Germany}}}

\affil[2]{\orgdiv{Institute of Molecular and Cell Biology}, \orgname{Mannheim University of Applied Sciences}, \orgaddress{\street{Paul-Wittsack-Straße 10}, \city{Mannheim}, \postcode{68163}, \country{Germany}}}

\affil[3]{\orgdiv{CeMOS}, \orgname{Mannheim University of Applied Sciences}, \orgaddress{\street{Paul-Wittsack-Straße 10}, \city{Mannheim}, \postcode{68163}, \country{Germany}}}

\affil[4]{\orgdiv{CHARISMA}, \orgname{Mannheim University of Applied Sciences}, \orgaddress{\street{Paul-Wittsack-Straße 10}, \city{Mannheim}, \postcode{68163}, \country{Germany}}}



\abstract{
    Biomedical research increasingly relies on 3D cell culture models and AI-based analysis can potentially facilitate a detailed and accurate feature extraction on a single-cell level. However, this requires for a precise segmentation of 3D cell datasets, which in turn demands high-quality ground truth for training. Manual annotation, the gold standard for ground truth data, is too time-consuming and thus not feasible for the generation of large 3D training datasets. To address this, we present a novel framework for generating 3D training data, which integrates biophysical modeling for realistic cell shape and alignment. Our approach allows the in silico generation of coherent membrane and nuclei signals, that enable the training of segmentation models utilizing both channels for improved performance. Furthermore, we present a new GAN training scheme that generates not only image data but also matching labels. Quantitative evaluation shows superior performance of biophysical motivated synthetic training data, even outperforming manual annotation and pretrained models. This underscores the potential of incorporating biophysical modeling for enhancing synthetic training data quality.
    }




\maketitle

\section{Main}\label{sec1}

    Biomedical and pharmaceutical research rely increasingly on three-dimensional cell culture models, such as spheroids and organoids. Leveraging optical tissue clearing alongside 3D-microscopy and AI-based image analysis enables the extraction of unprecedented detail at the single-cell level in 3D. As such, regional analysis of events like cell division and cell death can be performed. Most commonly, AI-based image analysis requires the manual generation of annotated datasets for model training, evaluation and subsequent model selection.
    To reach this surplus of information from 3D-cell culture whole mounts, the segmentation of the original 3D-datasets needs to be of maximal precision, which requests the availability of a sufficient amount of high-quality ground truth datasets. Additionally, for accurate segmentation of rare cell states, it is important that such objects are represented within the dataset. While manual expert annotation is often considered the gold standard for ground truth data, its labor-intensive nature, especially for voluminous 3D data, makes it impractical, usually taking long time periods to create a sufficient dataset. Therefore, new and realistic modes of generating 3D-ground-truth data are needed.
    \begin{figure}[b]
        \includegraphics[width=\linewidth]{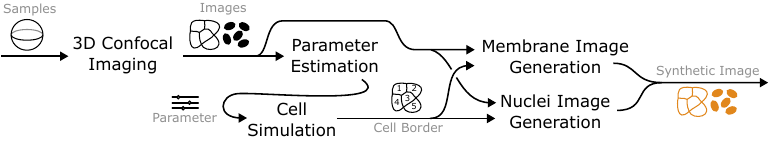}
        \caption{Presented workflow for generation of 3D synthetic images featuring nuclei and membrane signals. 3D cell cultures are imaged using confocal microscopy. Afterwards, parameters are extracted based on the recorded images which are then utilized during cell simulation to generate synthetic cell border images. Finally, synthetic nuclei and membrane images are generated on the basis of the simulated cell border image.}
        \label{fig:Overview_pipeline}
    \end{figure}
    
    Several methods have been proposed for the synthesis of 2D and 3D training data for nuclei~\cite{Fu2018, Dunn2019, Bruch2023, Yao2021, Wu2023, Eschweiler2021, Eschweiler2024} and cell~\cite{Eschweiler2021, Eschweiler2024} segmentation. We have recently presented SimOptiGAN, a method to create training data based on a combination of simulation and deep-learning-based optimization, allowing the introduction of rare elements. It generates data with a few real-world nuclei and their subsequent assembly in virtual spheroids, including the implementation of synthetic optical features, such as noise, signal loss in depth, and point-spread function before optimization~\cite{Bruch2023}. This gave surprisingly good results in terms of training efficacy. However, the method placed nuclei in a rather arbitrary manner, leaving doubts concerning the accuracy compared to the real-world distribution. For example, in cancer cell spheroids, dividing cells are often more concentrated towards the border of the 3D-culture, while necrotic or apoptotic cells are more abundant towards the spheroid center. Moreover, cells oftentimes exhibit a preference for tangential orientation relative to the outer hull~\cite{Desmaison2018}. Therefore, a consideration of biophysical features and real-world distribution of cells with specific features might further increase the fidelity of synthetic data sets. This enhancement could furthermore foster greater consistency between the synthetic and real-image domains, thereby mitigating the risk of unintentionally generating unwanted structures through deep-learning-based transformations introducing inconsistencies between image and label data~\cite{Eschweiler2021, Boehland2019}. However, correct physical modeling of cell structures is a difficult and complex task. To this date, no existing method incorporates physical modeling to enhance the organization of cell and nuclei structures.
    
    Several approaches to simulate 3D cell cultures in a biophysically realistic manner have been developed, each tailored to various scientific approaches for studying a broad spectrum of biological phenomena ~\cite{Amereh2021, Bowers2020, Ozik2018, Barham2023, Zhang2023}. 
    In particular, a widely used method to simulate the temporal evolution of cell morphology in 2D- or 3D-cell cultures is the Cellular Potts Model (CPM) \cite{Graner1992, Glazier1993}. Utilizing a grid-based system of pixels (2D) or voxels (3D), the CPM assigns each unit to specific cells or the extracellular medium.  By modeling the total free energy of the system and applying the Metropolis algorithm in a Monte Carlo simulation, the initial cell shapes evolve into their thermodynamic equilibrium \cite{Graner1992, Glazier1993, Swat.2012}. This technique is used in exploring a variety of biological processes, including viral infection \cite{FerrariGianlupi2022}, vascularization \cite{Shirinifard2012, Svoboda2016}, cell sorting and migration \cite{Libby2019, Mulberry2020, Guisoni2018, Niculescu2015, Scianna2021, Thomas2022, Tikka2022}, morphological changes \cite{Link2023, Hirway2021, Albert2014} and cancer research \cite{Li2014, Osborne2015, Szabo2013, Drasdo2005, Kim2013, Norton2018, Smallbone2007, Shirinifard2009, Rubenstein2008, Jeanquartier2016}. The application of the CPM or alternative biophysically driven models to replicate the morphology of cell cultures for the creation of synthetic training data, as detailed in this study, has not yet been reported in scientific literature.

    To improve the cell arrangement in synthetic image data, we introduce a framework to generate 3D training data of nuclei and membrane signals, integrating biophysical modeling to achieve realistic cell alignment and orientation (see \autoref{fig:Overview_pipeline}). Unlike existing approaches, our framework facilitates the creation of coherent membrane and nuclei signals, enabling training of segmentation models utilizing both signals simultaneously for improved segmentation performance. Three novel approaches for generating nuclei signals (SimOptiGAN+, Mem2NucGAN-P, Mem2NucGAN-U) are presented and compared against a previously developed synthesis method SimOptiGAN. The random nuclei placement process of SimOptiGAN is improved in SimOptiGAN+ by utilizing simulated cell borders to control the nuclei placement process. Mem2NucGAN-P and Mem2NucGAN-U utilize generative adversarial networks (GANs) to generate nuclei signals based on the simulated cell borders. For this, a new training and post-processing scheme is introduced, for the first time allowing the additional generation of nuclei labels corresponding to the generated signal images. This adaptation can also be utilized for other image modalities. For Mem2NucGAN-P the training is performed with paired images while Mem2NucGAN-U utilizes unpaired images. We furthermore prove superior performance of synthetic nuclei training data against manually annotated data and a pretrained Cellpose model, using three manually corrected ground truth patches for comparison. This is also the first time, synthetic training data of SimOptiGAN is evaluated on fully annotated image data.

\section{Results}
    
    \subsection{Biophysical modeling enables marker transformation}
\begin{figure}
            \centering            
            \includegraphics[width=1\linewidth]{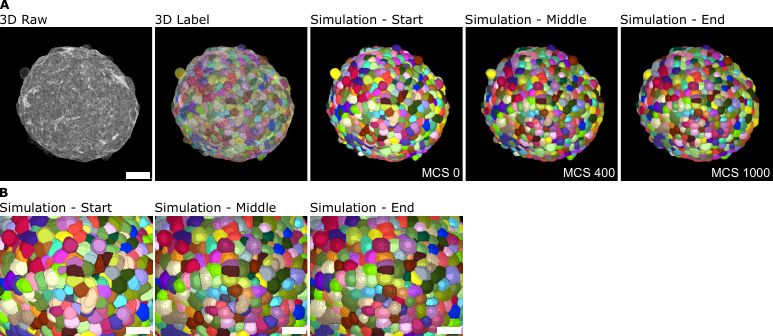}
            \caption{Biophysically-driven 3D spheroid simulations from real image data reveal cell morphology variability across Monte Carlo steps (MCS), while preserving the distribution of morphological features throughout the spheroid. \textbf{A} 3D maximum projection of membrane marker SiR-actin (\textbf{grey,far left}), overlay with label masks (\textbf{second from the left}), and snapshots of the corresponding CPM simulation of morphological changes at the start (\textbf{middle}, 0 MCS), middle (\textbf{second from the right}, 400 MCS) and end (\textbf{far right}, 1000 MCS) of the simulation. \textbf{B} Zoom images of the simulation steps shown in (a). Scale bars: (a) \SI{25}{\micro\metre}, (b) \SI{15}{\micro\metre}.}             
            \label{fig:Biophysical_Simulation}
        \end{figure}

        To integrate meaningful, biophysical information into the generation of synthetic images, CPM simulations were developed using 3D label masks obtained from automated whole mount segmentation of real image data (\autoref{fig:Biophysical_Simulation}). The detection of membranes relied on minimal training on single optical planes, acknowledging a balance between invested time and accepting some inaccuracies in boundary detection. Subsequently, morphological features such as volume, surface area, and shape descriptors were extracted from individual labels to analyze the distribution of cellular characteristics across the spheroid. Label masks obtained from whole mount spheroids were then transformed into a starting configuration for a CPM simulation. Furthermore, each cell's target volume and surface area were preset to their individual starting values.

        Before simulating the entire spheroid, a parameter estimation was conducted on a small subsection to identify optimal combinations of contact energies $J^\textrm{(c-c)}$ and $J^\textrm{(c-m)}$, as well as the volume and surface constraints $\lambda_V$ and $\lambda_A$. These parameters influenced cellular morphology, enabling changes in shape and position while preserving the overall distribution of features within the simulated spheroid. Evaluation of individual parameter combinations was performed by minimizing a metric $m$ that compares the start and endpoint of the simulation by using the Wasserstein distance of the morphological feature distribution $W_k$ and the mean intersection over union (IoU) of all cells, as described in Section \ref{Sec:biophys_model}.
       
\begin{figure}
            \centering            
            \includegraphics[width=0.8\linewidth]{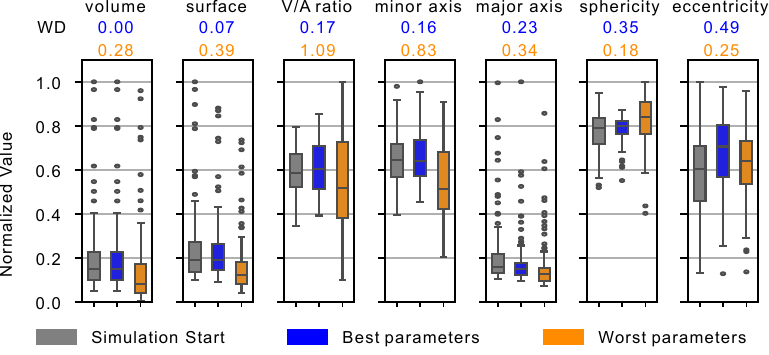}
            \caption{Comparison of morphological feature distributions at the start and end of simulations reflecting changes in cell morphology, for the best and worst parameter sets according to the metric $m$, Eq.\,\eqref{eq:metric_paramscan}. Data is derived from a manually segmented image patch. Individual Wasserstein distances, quantifying the variations between the simulation's start and end for each feature, are marked above the plots for the best (blue, $\lambda_V=10.0$, $\lambda_A = 0.001$, $J^\textrm{(c-c)} = 2.0$, $J^\textrm{(c-m)} = 55.0$) and worst (orange, $\lambda_V=0.001$, $\lambda_A = 10.0$, $J^\textrm{(c-c)} = 10.0$, $J^\textrm{(c-m)} = 10.0$) parameter sets. Boxes indicate quartiles of data, while whiskers encompass all values within 1.5 times of the IQR, and outliers are indicated as single data points.}             
            \label{fig:Parameter_Scans}
        \end{figure}

       \autoref{fig:Parameter_Scans} displays the distribution of selected morphological features across individual cells at the start of the simulation, based on real-world data. It also presents results for these features after 1000~Monte-Carlo-Steps (MCS), comparing two sets of parameters that received the highest and lowest rankings in the parameter scans. 
        
        After parameter estimation, sets of parameters resulting in the smallest values of metric $m$ were selected for simulations of whole spheroids. These simulations were then conducted for 1,000~MCS. As a result of the simulations, 3D cell border images were received, where each cell is described as a set of voxels with the same intensity. Representative 3D images of an HT-29 spheroid, including fluorescent membrane markers and a segmented label mask, along with the corresponding 3D simulation images from the start (0~MCS), middle (400~MCS), and end (1000~MCS) of the simulation, are displayed in \autoref{fig:Biophysical_Simulation}a. Zoom images from the respective time points of the simulation highlight the morphological changes of individual cells throughout the simulation.

    \subsection{Synthetic data matches structure of real counterparts}
\begin{figure}
            \centering
            \includegraphics[width=\linewidth]{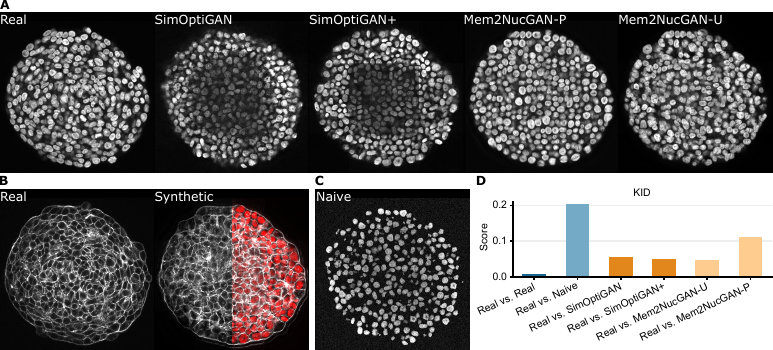}
            \caption{Comparison of real and synthetic images. \textbf{A} shows image slices of real and synthetic 3D nuclei images. SimOptiGAN uses a random process for nuclei arrangement, while SimOptiGAN+, Mem2NucGAN-P and Mem2NucGAN-U incorporate biophysical modeling for a realistic arrangement. \textbf{B} shows image slices of real and synthetic membrane signals. The synthetic membrane signal is generated based on the same simulated cell borders used in the nuclei synthesis methods SimOptiGAN+, Mem2NucGAN-P and Mem2NucGAN-U. Consequently, the membrane signal exhibited a consistent cell arrangement, as demonstrated by the overlay of synthetic membrane with nuclei generated using SimOptiGAN+. \textbf{C} shows a preview of naively generated data used as worst example for the KID evaluation in D. \textbf{D} comparison of synthetic nuclei images with real counterparts based on the Kernel Inception Distance (KID). Lower scores represent a greater similarity between real and synthetic signals.}
            \label{fig:img_comp}
        \end{figure}

        A human approach for the quality evaluation of synthetic image data is the visual comparison to their real counterpart. However, the visual differences a human perceives may differ from the ones that are of importance for a segmentation model. The Kernel Inception Distance (KID) \cite{Binkowski2018} enables the comparison of images in the feature space of a network and thus focuses on differences more relevant for the segmentation task. Lower distance values represent a closer similarity of real and synthetic images, and therefore a better result.
        \autoref{fig:img_comp}a shows 2D samples of synthetic image data generated with the newly introduced methods SimOptiGAN+, Mem2NucGAN-P and Mem2NucGAN-U and a previous method SimOptiGAN without any biophysical modeling in comparison with a real counterpart. Additionally, the KID between generated 3D images and real counterparts is given (\autoref{fig:img_comp}d). As comparison, the KID is also given between real and real samples, indicating the best possible result and real and naively generated 3D images (\autoref{fig:img_comp}c), indicating a suboptimal result.
        
        By visual comparison, the synthesis methods showed differences in nuclei morphology and arrangement, as well as differences in brightness and texture. The nuclei on the data generated with SimOptiGAN, i.e., without biophysical modeling,  appeared less sharp, and featured a more uniform texture. By incorporating biophysical modeling (SimOptiGAN+) the nuclei arrangement became more structured, e.g., with small holes in the spheroid. Both of the images, i.e., of SimOptiGAN and SimOptiGAN+, also exhibited a reduced brightness in the center region. Data generated with Mem2NucGAN-P displayed rather large and more roundish nuclei. Furthermore, a light checkerboard texture pattern was visible. Data generated with Mem2NucGAN-U led to the smallest KID when comparing to real data. However, some nuclei morphed into each other, showing no clear outline between them.
        
        \autoref{fig:img_comp}b shows a synthetic membrane generated on the same biophysical simulated cell borders as the nuclei images generated with SimOptiGAN+, Mem2NucGAN-P and Mem2NucGAN-U (\autoref{fig:img_comp}a). Both signals feature the same cell arrangement and can be combined. This is illustrated by overlaying synthetic membrane signals with nuclei signals generated using SimOptiGAN+ (\autoref{fig:img_comp}b).
        
        The comparison of the generated images with the KID measure (\autoref{fig:img_comp}d) shows only small differences between SimOptiGAN, SimOptiGAN+ and Mem2NucGAN-U compared to the difference to Mem2NucGAN-P. All methods, however, show drastically lower KID scores compared to a naive approach. Apart from the central dark regions obtained in the nuclei images by SimOptiGAN and SimOptiGAN+, the KID measure correlated apparently well with the visual impression.

    \subsection{Synthetic data can outperform manual annotation and universal models in terms of segmentation performance}

\begin{figure}[!pht]
                \includegraphics[width=\linewidth]{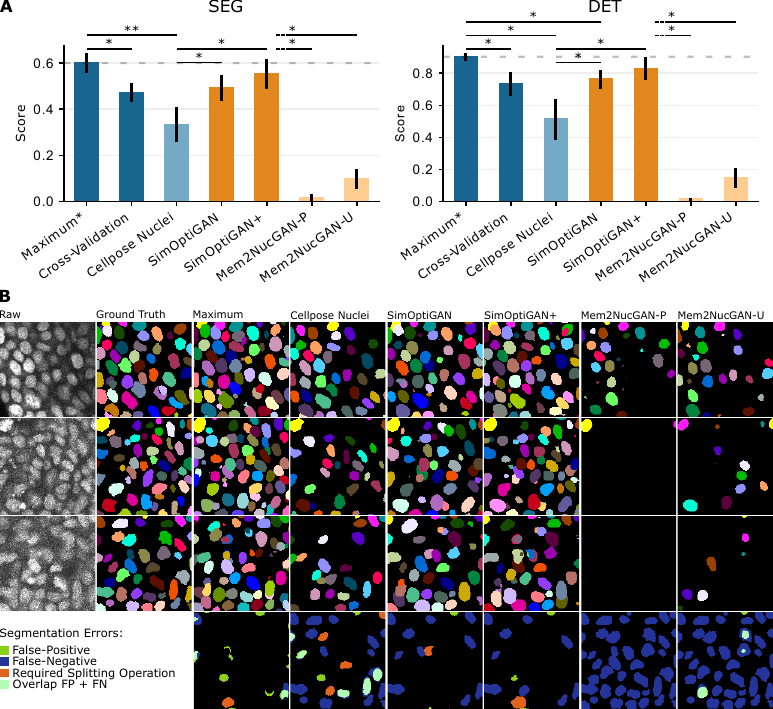}
                \caption{\textbf{A} SEG and DET segmentation scores of nuclei segmentation models trained with different type of training data. Scores can range from zero (worst-possible) to one (best-possible). Three manual corrected image patches of different image regions serve as test data. Maximum indicates a model that was trained on the test data and is considered an upper boundary. Dark blue color indicates training data generated by manual annotation. The nuclei model provided by Cellpose is depicted in light blue color. Dark and light orange colors indicate the pure use of synthetic training data generated with physical simulation based and GAN-based approaches, respectively. The error bars represent the standard deviation of model performance across the three ground truth patches.
                \textbf{B} qualitative comparison of nuclei segmentation results. Representative single optical sections of ground truth patches are shown for enhanced clarity. The first and second column display the raw image signal and its corresponding ground truth, while subsequent columns show the segmentation masks obtained with the segmentation models. Additionally, the last row visualizes the DET related errors of the third row, including false-negative, false-positive, and required splitting operations. A complete visualization of DET errors across all ground truth patches is given in the supplementary material (Fig.~A2).}
                \label{fig:ctc_res}
            \end{figure}

            The segmentation performance of the introduced methods was evaluated using the SEG (segmentation) and DET (detection) metrics as utilized in the Cell Tracking Challenge~\cite{Ulman2017}. While the SEG metric focuses on the overlap between segmentation and ground truth, the DET metric evaluates the results on a cell level, i.e., whether an object is correctly detected or not. To compare the segmentation performance of different training datasets, the 3D StarDist \cite{Weigert2020} segmentation model was used. The performance was calculated based on three ground truth images of size \qtyproduct[product-units = power]{25 x 128 x 128}{\px} to \qtyproduct[product-units = power]{51 x 128 x 128}{\px} containing in total $1001$ nuclei. Ca. $\SI{37.5}{\hour}$ were required for ground truth generation.
            \autoref{fig:ctc_res}a shows the obtained segmentation performances for the introduced training data generation methods.

            The model labeled as Maximum was trained on the ground truth dataset later used for testing. This model was considered as a maximum of the segmentation performance for the StarDist architecture on this dataset, as train and test data were identical. However, generating training data through manual annotation of 3D image data is a challenging and time-consuming process. Cross-Validation indicates results obtained by using a leave-one-out approach on the ground truth dataset, i.e., using each patch as the test set once while training the model on the remaining two patches. For inference of the Cellpose nuclei model~\cite{stringer2021cellpose}, only the nuclei size parameter was adapted to match the size of the ground truth data. The remaining models were all trained with synthetic datasets. SimOptiGAN was generated using our preliminary synthesis method presented in \cite{Bruch2023} which does not incorporate biophysical modeling. Conversely, the datasets labeled SimOptiGAN+, Mem2NucGAN-P and Mem2NucGAN-U incorporate biophysical modeling.
            
            The results revealed significant variations in the segmentation performance among the different models. The maximum SEG score for this data was $0.6$ (Maximum). Using Cross-Validation the score dropped to $0.472$. Cellpose Nuclei attained a SEG score of $0.332$. The SimOptiGAN model yielded a SEG score of $0.492$ which is improved to $0.553$ - the best score apart Maximum - by incorporating biophysical modeling (SimOptiGAN+). Both, Mem2NucGAN-P and Mem2NucGAN-U demonstrated notably lower SEG scores, with values of $0.017$ and $0.098$, respectively.
    
            The DET scores across all models exhibited a slight increase compared to the SEG scores, given that only detection errors are penalized. Nonetheless, the disparity between the models remained consistent. The Maximum and Cross-Validation scores for the DET metric were $0.903$ and $0.733$, respectively. The Cellpose Nuclei model attained a DET score of $0.512$. SimOptiGAN+ again, improved the result from $0.763$ (SimOptiGAN) to $0.829$. Both Mem2NucGAN-P and Mem2NucGAN-U led to scores of $0.013$ and $0.146$, respectively.
    
            \autoref{fig:ctc_res}b shows the segmentation results in comparison with the raw image and the ground truth obtained by manual annotation. Shown is a 2D slice for each of the three ground truth patches. The patch locations were selected to feature three image regions: the upper, middle and lower region of the 3D image stack. As a result, the brightness and signal-to-noise ratio were decreasing over the three patches. At first glance, it can be seen that the models Cellpose Nuclei, Mem2NucGAN-P and Mem2NucGAN-U had difficulties detecting nuclei in darker image regions. With the segmentation model, Mem2NucGAN-P only two nuclei were detected in the slice of the second patch and none in the slice of the third patch. This matches with the results of DET measure. The model Maximum detected most of the nuclei, directly followed by the SimOptiGAN+ and SimOptiGAN model.
    
    \subsection{Using synthetic data drastically reduces manual effort for training data generation}
        By employing the presented framework for synthetic data generation, users can drastically reduce the manual effort required to generate a sufficient amount of training data for 3D segmentation models. While the time effort for manual annotation is reduced, computation time and GPU hardware are required. A comparison of the times needed for synthetic training data generation with the proposed methods is given in \autoref{tab:time_comp}.
        \begin{table}
            \centering
            \begin{tabular}{lrrr}
                Method            & Manual annotation & Computation preparation        & Computation inference     \\ \hline
                Maximum           & \qty{37.5}{\hour} & -                  & -                   \\
                SimOptiGAN        & \qty{0.4}{\hour}  & \qty{99.5}{\hour}  & $<$\qty{0.1}{\hour} \\
                SimOptiGAN+       & \qty{4.4}{\hour}  & \qty{112.5}{\hour} & \qty{148}{\hour}    \\
                Mem2NucGAN-P      & \qty{4}{\hour}    & \qty{33.5}{\hour}  & \qty{148}{\hour}    \\
                Mem2NucGAN-U      & \qty{4}{\hour}    & \qty{72.5}{\hour}  & \qty{148}{\hour}    \\
            \end{tabular}
            \caption{Comparison of required times to generate training data. The Cellpose nuclei segmentation model is not included, as no training data is required for the usage of this model. Manual annotation defines the time required to manually generate or correct label masks. Preparation and inference time include the computation time required for the generation of synthetic images. While preparation contains one-time calculations such as parameter estimation for biophysical simulation and GAN-based model training, data generation consists of the remaining calculation time to generate four synthetic images.}
            \label{tab:time_comp}
        \end{table}

        The biophysical simulation used to create realistic cell borders involved a one-time segmentation of membrane signals to automatically extract essential parameters. For this, the pretrained Cellpose Cyto2 model was optimized using the human-in-the-loop annotation method, requiring 4 h of annotation time. Subsequent parameter estimation for biophysical simulation needed \qty{13}{\hour} of calculation time. Both the membrane segmentation and simulation parameter estimation were one-time processes. After these initial steps, the user can theoretically generate an infinite number of simulated cell borders. The computation for the biophysical simulation took \qty{148}{\hour} to produce four 3D cell border images, which served as the basis for generating realistic nuclei and membrane signals. As SimOptiGAN does not utilize biophysical simulated cell borders, the associated annotation and calculation times were not applicable.
        
        For generating nuclei signals with SimOptiGAN and SimOptiGAN+, manual annotation of a few nuclei was necessary to populate the nuclei prototype database. In this study, only $10$ nuclei were annotated as prototypes, requiring $\SI{22.5}{\min}$ for annotation. The most resources for SimOptiGAN and SimOptiGAN+ were needed for training the optimization network ($\qty{99.5}{\hour}$ and $\qty{42}{\giga\byte}$ of GPU memory). For the other parts of the SimOptiGAN and SimOptiGAN+ pipeline (prototype generation and imaging simulation), only $\qty{0.5}{\giga\byte}$ of RAM and a computation time of less than two minutes were required. In contrast to previous experiments \cite{Bruch2023}, calculations of SimOptiGAN (and SimOptiGAN+) were directly performed at the desired output resolution, which drastically improved calculation time and memory consumption.
        
        Mem2NucGAN-P and Mem2NucGAN-U did not necessitate additional manual annotation but also needed training of GAN-based models. Training of the transformation network Mem2NucGAN-P required \qty{20.5}{\hour} and \qty{16.5}{\giga\byte} GPU memory, while Mem2NucGAN-U consumed \qty{59.5}{\hour} and \qty{31}{\giga\byte} GPU memory. 

        Initial membrane segmentation and  subsequent parameter estimation and biophysical simulation were performed on a workstation equipped with an AMD Ryzen 9 5950X CPU, 128 GB RAM, and an NVIDIA GeForce RTX 3060 Ti 8~GB GPU. All remaining calculations were performed on a server computer equipped with an AMD EPYC 7252 CPU, 64 GB RAM, and an NVIDIA A6000 48~GB GPU.

\section{Discussion}
    For the first time, biophysical simulation was here incorporated into the generation process of synthetic membrane and nuclei signals. Both synthetic signals were based on the same simulated cell borders and therefore allowed the combination of both signals. This was utilized to train segmentation models with both membrane and nuclei signals as input which, as preliminary results showed, improved segmentation performance for membrane signals.

    We furthermore introduced a new GAN training scheme which enables the generator to not only transform image signals, but more importantly, also generate the corresponding labels. This, for the first time, allowed to use GANs to generate nuclei signals and their labels based on membrane signals, which are afterwards available as training data for segmentation models. This approach could potentially be used in the inverse direction, i.e., for the generation of membrane signals and labels based on real nuclei signals or even for other image modalities like the transformation between nuclei and the proliferation marker Ki-67.

    The incorporation of biophysical modeling visually improved the arrangement of nuclei compared to a random placement approach.
    Visual inspection of the synthetic images generated by various methods (see \autoref{fig:img_comp}) revealed notable differences in the brightness distribution. Specifically, images produced using the SimOptiGAN and SimOptiGAN+ showed a decreasing brightness deeper inside the spheroid, whereas those generated by the Mem2NucGAN methods maintained a nearly constant brightness throughout the image.
    One possible explanation for this discrepancy could stem from the utilization of different training datasets for the optimization model used in SimOptiGAN(+) and the transformation models used in Mem2NucGAN. The training data for the latter models consisted of real images exhibiting only a minimal reduction in brightness in deeper regions. Additionally, the absence of brightness information in simulated binary membranes, coupled to a patch-based transformation approach, rendered the Mem2NucGAN models mostly incapable of incorporating a decreasing brightness in deeper regions.
    
    Such discrepancies undoubtedly affected the segmentation quality of nuclei in darker image regions. As observed in the visual inspection of the segmentation results (see \autoref{fig:ctc_res}), both Mem2NucGAN models encountered challenges in detecting nuclei in darker image regions. Both of the SimOptiGAN methods performed far better under such conditions, indicating that the brightness reduction was improving the segmentation performance.

    Another factor contributing to the lower segmentation scores of the Mem2NucGAN models could be attributed to the label generation process. As the binary labels were only an additional output of the generator models, a strict alignment between image and label data may not be given for all instances. Although the post-processing step, used to derive instance labels, visually improved the quality of the label data, it may have introduced additional discrepancies between labels and generated image data.

    The additional membrane segmentation step required for Mem2NucGAN-P might have affected the task of transformation learning due to errors in the  membrane segmentation. We also tested additional approaches which did not require segmented membrane signals for training of the generator and instead directly use the raw membrane signals (see supplementary material Fig.~A1, \emph{Real Membrane $\to$ Nuclei}). However, these involved a pre-processing of the simulated binary membrane to match real membrane signals. We tested several pre-processing variants including the imaging-simulation and a GAN transformation model to generate realistic membrane signals, however, no nuclei with realistic textures could be obtained.

    The segmentation scores showed, that synthetic training data can not only outperform the pretrained Cellpose model, but also training data generated by manual annotation. It can be argued, that by increasing the data size of the manual training data the results of the Cross-Validation model would be improved, this would also strongly increase the amount of time required for annotation. Ideally, smaller patch sizes are preferred to label a greater number of patches, thereby enhancing heterogeneity. However, the requirement for a minimal patch size in most segmentation models restricts the number of patches that can be labeled within a limited timeframe. In the case of rare objects, a reduced number of patches can lead to an underrepresentation, or even a complete absence of such objects in the training dataset. Synthetic training data, on the other hand, can be generated in large quantities with minimal user interaction and thus with minimal time investment. SimOptiGAN and SimOptiGAN+ additionally allow for an overrepresentation of rare objects and thus improve segmentation performance for these objects. Both, SimOptiGAN and SimOptiGAN+, also require some annotation for the extraction of nuclei prototypes. However, the time investment here is rather small. For the generated synthetic data, only 20 nuclei were labeled instead of $1001$ nuclei present in the three ground truth patches.

    No clear statement can be given as to why the incorporated biophysics improved the results of the segmentation. Apart from the placement process of the nuclei, the pipelines were identical, including the used parameters. E.g., the same nuclei prototypes, imaging-simulation parameters and even the same optimization model were used. Reasonable explanations for the improvement may be due to a more fitting nuclei size or a more realistic orientation and arrangement of nuclei. Both could potentially lead to a better transformation during the optimizing step. 

\section{Conclusion}

    In contrast to existing approaches, our three new approaches incorporated biophysical simulation to improve the fuzzy similarity between real-world and synthetic data, thus also enhancing training quality of segmentation algorithms.
    Furthermore, the segmentation model trained with our synthetic data outperformed both the pretrained Cellpose Nuclei model and a model trained with manually annotated data. This use of synthetic training data enabled precise single-cell level analysis of 3D cell cultures. Looking ahead, we aim to test our GAN-based methods, Mem2NucGAN-P and Mem2NucGAN-U on other image modalities, e.g., nuclei and Ki-67, and assess the segmentation performance of models trained with combined synthetic nuclei and membrane signals.

\section{Methods}
        
    \subsection{Biophysically motivated simulation of cell boundaries}\label{Sec:biophys_model}
        Cell boundaries were generated in a biophysically motivated simulation. To this end, a 3D Cellular Potts Model (CPM) was set up, using the \textit{CompuCell3D} implementation \cite{Swat.2012} (Version 4.3.2, Revision 0). The simulation approach outlined below is described in more detail in \cite{Nuerberg2024_preprint}.

        As initial configuration of the simulation, we started from a real-world confocal image stack of a mono-culture spheroid with fluorescence labeled cell membranes, as described below in Section \ref{Sec:Dataset}.
        
        Cell boundaries in this image stack were segmented with \textit{Cellpose 2.0} \cite{stringer2021cellpose}, using a custom model based on the pre-trained \textit{cyto2} model \cite{pachitariu2022cellpose, Nuerberg2024_preprint}.
        The resulting 3D label image was then further processed: Label masks with less than 5 voxels were removed, as they result almost certainly from segmentation errors. Labels spanning less than four image planes in z-direction were removed for the same reason. A closing and a dilation operation was applied to fill the resulting holes in the image. Image up-sampling through nearest neighbor interpolation was applied in z-direction to ensure isotropic voxel size of \SI{0.5682}{\micro m^3}. A list of morphological features $F_k$ was extracted for each cell, where the index $k$ runs over 7 distinct features: cell volume, surface area, volume-to-surface (V/A) ratio, minor and major axis length, sphericity and eccentricity.
        
        The label image was then converted into a file format that specifies the initial configuration of a CPM simulation in CompuCell3D \cite{Nuerberg2024_preprint}.
        The system Hamiltonian $H$, which models the total free energy of the cell conglomerate, included terms for the adhesion energy between neighboring cells, surface free energy contributions of cells in contact with the medium, and penalty terms for the deviation of volume $V_i$ and surface area $A_i$ of cell $i$ from target values:

        \begin{align}
            H =& \sum_i \lambda_V \left(V_i - V^\textrm{(target)}_{i}\right)^2 + \sum_i \lambda_A \left(A_i - A^\textrm{(target)}_{i}\right)^2 \\
            &+\sum_{i,j} J^\textrm{(c-c)} A^\textrm{(c-c)}_{i,j} + \sum_{i} J^\textrm{(c-m)} A^\textrm{(c-m)}_i
        \end{align}

        Here, $\lambda_V$ and $\lambda_A$ are penalty parameters which determine the strength of the volume / area constraints. $A^\textrm{(c-c)}_{i,j}$ is the contact area between cell $i$ and $j$, $A^\textrm{(c-m)}_i$ the contact area of cell $i$ with the surrounding medium, and $J^\textrm{(c-c)}$ and $J^\textrm{(c-m)}$ are the respective surface tension parameters.
        The target values $V^\textrm{(target)}_{i}$ and $A^\textrm{(target)}_{i}$ for volume and area were assigned individually to each cell. Here, the target values of cell $i$ were taken from the \textit{actual} values of cell $j$ in the initial configuration, using a random permutation $P_{ij}$ of all cells. This way, each cell has target values \textit{different} from its initial values, ensuring some dynamics in the simulation; on the other hand, the distribution of target values and original values over all cells is identical, ensuring a realistic volume/area distribution of the ensemble throughout the simulation.

        Moreover, a temperature parameter $T$ needs to be set, 
        which determines the strength of the fluctuations around the energy minimum in the equilibrium state.
        
        In sum, the simulation model contained 4 parameters: $\lambda_V, \lambda_A, J^\textrm{(c-c)}, J^\textrm{(c-m)}$. Their values were fixed by a parameter scan on a hand-segmented subset of a spheroid of \qtyproduct[product-units = power]{101x 250 x 250}{\px} (z, y, x), with the objective to minimize the following metric $m$ \cite{Nuerberg2024_preprint}:
        \begin{equation}
            m = \left(\frac{1}{N_\textrm{Features}}\sum_k^\textrm{Features} W_k \right) \cdot \left(\frac{1}{N_\textrm{Cells}}\sum_i^\textrm{Cells} \mathrm{IoU}_i \right)
            \label{eq:metric_paramscan}
        \end{equation}
        
        Here, $W_k$ is the Wasserstein distance of the empirical distribution of the morphological feature $F_k$ at the beginning of the simulation vs. its end.
        Small values of $W_k$ thus ensure that throughout the simulation, the cell ensemble retains its morphological characteristics.
        $\mathrm{IoU}_i$, on the other hand, is the intersection over union of cell $i$ in the beginning vs. the end of the simulation; i.e., it measures to what extent a cell moves in the simulation. Small values of $\mathrm{IoU}_i$ hence ensure that individual cells change their position and/or shape in the simulation, to avoid simulation results with little movement or even a complete 'freeze'. The resulting parameter values are listed in \autoref{table:ParamScanValues}.

        \begin{table}[htb]
            \centering
            \caption{Simulation parameters for CompuCell3D.}
            \label{table:example_label}
            \begin{tabular}{l c r}
                Parameter & Parameter Scan Values & Best \\ \hline
                $\lambda_V$ & $0.001$, $2.0$, $4.0$, $6.0$, $8.0$, $10.0$ & $10.0$ \\
                $\lambda_A$ & $0.001$, $2.0$, $4.0$, $6.0$, $8.0$, $10.0$ & $0.001$ \\
                $J^\textrm{(c-c)}$ & $0.0001$, $2.0$, $4.0$, $6.0$, $8.0$, $10.0$ & $2.0$\\ 
                $J^\textrm{(c-m)}$ & $10.0$, $55.0$, $100.0$ & $55.0$ \\ 
                $T$ & -- & $30.0$ \\
                Potts Neighbor Order & -- & $3$ \\
                Contact Neighbor Order & -- & $4$ \\
            \end{tabular}
            \label{table:ParamScanValues}
        \end{table}

        With those parameters, the system was simulated over 1000 MCS with different random seeds in CompuCell3D. The number of MCS was selected because, beyond this point, the system approaches thermodynamic equilibrium, resulting in only minimal changes in cellular morphology. The resulting label masks were converted into cell border images, as a template for the actual generation of synthetic images, as described below.
        
    \subsection{Membrane synthesis}
        To produce authentic membrane signals derived from the simulated binary membranes, we employed the CycleGAN-architecture\cite{Zhu2017}. Apart from the 3D adaptation of the generator and discriminator models, the CycleGAN-architecture is similar to the original one presented in \cite{Zhu2017}. As training data, unpaired image data consisting of four images showing real membrane signals and four images showing simulated binary membranes is utilized.
        
    \subsection{Nuclei synthesis - from membrane to nuclei}
        The main objective was the generation of synthetic nuclei data for the training of segmentation models. We analyzed several ways of creating nuclei images based on the simulated cell borders (see~\autoref{fig:method_overview} and a comprehensive overview is given in the supplementary material Fig.~A1). One approach is based on placing nuclei into the simulated cell borders, while the other two approaches utilize a GAN-based transformation between membrane like signals and nuclei signals. These two approaches are subdivided based on the model which enables the transformation.
        \begin{figure}[hb]
            \centering
            \includegraphics[width=1\linewidth]{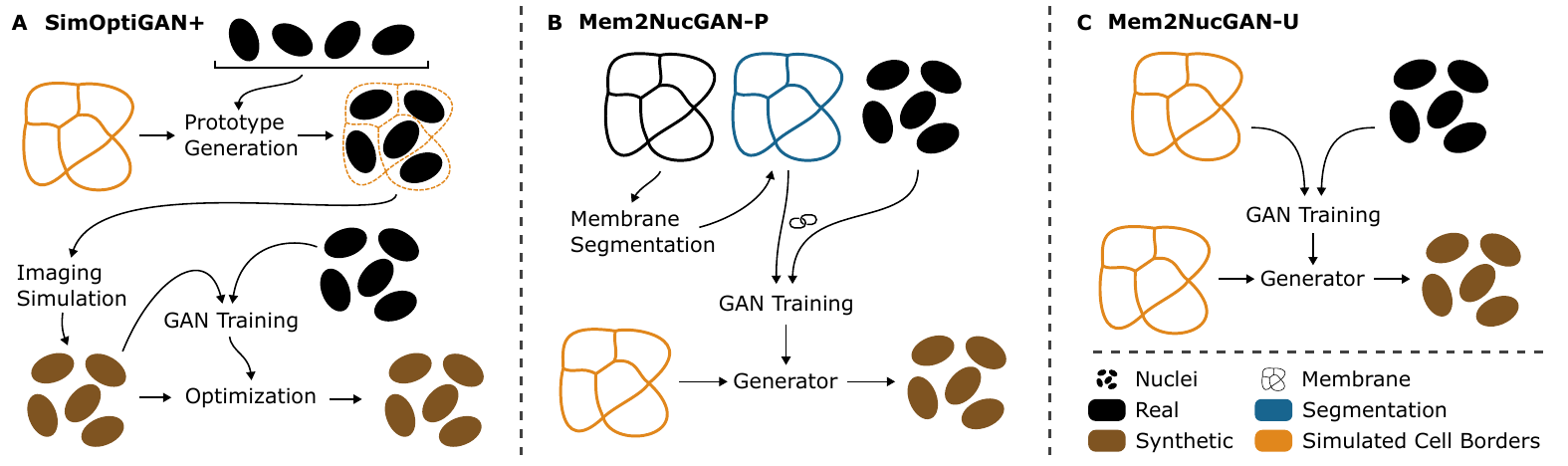}
            \caption{Overview of presented methods for the generation of nuclei images based on simulated cell labels. Mem2NucGAN-P and Mem2NucGAN-U rely on a GAN-based transformation, while SimOptiGAN+ uses the simulated cell borders to improve nuclei placement of a previously presented simulation-based pipeline.}
            \label{fig:method_overview}
        \end{figure}

        \paragraph{Simulation based approach with random nuclei placement: SimOptiGAN}
            Our pipeline firstly introduced in \cite{Bruch2023} enables the generation of realistic looking 3D nuclei images. It relies on a database containing nuclei prototypes, which are extracted by manual annotation of a few nuclei in high-resolution images. During the \emph{Prototype Generation}, a 3D image of a cell culture is generated by placing nuclei prototypes into the image. Afterwards, an \emph{Imaging Simulation} is performed to simulate the effects of the recording process with a microscope. Lastly, during the \emph{Optimization}, a GAN is used to post-process the generated image.
        
        \paragraph{Enhanced nuclei placement for simulation-based approach: SimOptiGAN+}    
            With SimOptiGAN, the arrangement of nuclei is not physically motivated, but rather follows a random pattern. For SimOptiGAN+, the pipeline is adapted to utilize the cell borders generated by biophysical simulation to improve the placement process of the nuclei prototypes during the \emph{Prototype Generation} step (see~\autoref{fig:method_overview}a). The new procedure is described in the following.
            
            For each cell of the simulated cell structure, a random nucleus prototype is chosen from the database. This prototype is then rotated to align with the cell's orientation and scaled to match a predefined volume relative to the cell's volume. As nuclei in the real world are not strictly located in the center of a cell, the placement position is selected based on a random distribution with its highest likelihood in the cell's center. For a selected position, the overlap between the nucleus prototype and cell is assessed. If it is larger than a specified percentage threshold of the nucleus prototype's volume, the nucleus is placed. If the overlap is smaller than the threshold, a new position is selected and reassessed. This process continues until either a suitable position is found or the maximum number of attempts is reached. In the later case, a new nuclei prototype is chosen from the database and the positioning procedure is repeated. A maximum number of nuclei prototypes are tested, before the cell is skipped and no nucleus is placed in this cell.

            Once all cells are processed, the \emph{Prototype Generation} is completed. Afterwards, the \emph{Imaging Simulation} is used to simulate the optical and sensor effects of a microscope. This includes a brightness reduction in deeper regions, the convolution with a point spread function, a downsampling step and the calculation and application of noise. Subsequently, a deep-learning-based post-processing is performed during the \emph{Optimization} step, to further increase the realness of the image. For this step, the same network can be used as for the regular pipeline. Finally, a synthetic nuclei image is obtained with the corresponding labels.

        \paragraph{Binary membrane to nuclei with paired data: Mem2NucGAN-P}
            This approach utilizes a GAN-based transformation for the generation of realistic nuclei signals based on the biophysical simulated cell boundary images (see \autoref{fig:method_overview}b). Our previous experiments indicated that GANs can learn the generation of nuclei signals with correct orientation and morphology by only utilizing membrane signals. Using such a transformation model which are trained on real data, would thus allow the generation of nuclei signals with realistic properties corresponding to the membrane signal.

            However, as the biophysical simulated cell borders are binary signals, the transformation will fail due to the domain gap to the initial training data (real membrane signals). To overcome this, we train the transformation model with segmented instead of raw membrane signals.
            This requires the training of a \emph{Membrane Segmentation} model. As manual annotation of membrane images is not feasible, synthetic training data is generated by an additional transformation model \emph{Synthetic Cell Borders $\to$ Membrane} (see supplementary material Fig.A1). This CycleGAN based transformation model allows the generation of realistic membrane signals based on the biophysical simulated cell borders by utilizing unpaired images as training data. 
            Data pairs of biophysical simulated cell borders and synthetic membrane signals generated with \emph{Synthetic Cell Borders $\to$ Membrane} are then used to train the \emph{Membrane Segmentation} model.

            By utilizing the \emph{Membrane Segmentation} model, real image pairs of segmented membrane and corresponding raw nuclei signals can be generated (see \autoref{fig:method_overview}b). This data is then used to train a conditional GAN (cGAN) representing the main transformation model \emph{Binary Membrane $\to$ Nuclei}. Finally, this main transformation model is used to generate synthetic nuclei signals based on biophysical simulated cell borders.
            
        \paragraph{Binary membrane to nuclei with unpaired data: Mem2NucGAN-U}
            While the transformation model of the previous approach is trained on real data, and therefore can learn morphological correlations between membrane and nuclei, it is a complex procedure. The required segmentation model and its training introduce additional error sources. Furthermore, differences between segmented membranes used during training and simulated membrane labels used during the inference could decrease the quality of the generated nuclei signals. We therefore tested an additional method which is based on a CycleGAN and thus does not require paired data for training (see~\autoref{fig:method_overview}c). The additional segmentation of membrane signals is skipped, and simulated membrane labels in combination with real nuclei signals are directly used for training. During inference, the model transforms the same input data as used during training, i.e., simulated membrane labels, into synthetic nuclei images.

\begin{figure}
                \centering
                \includegraphics[width=1\linewidth]{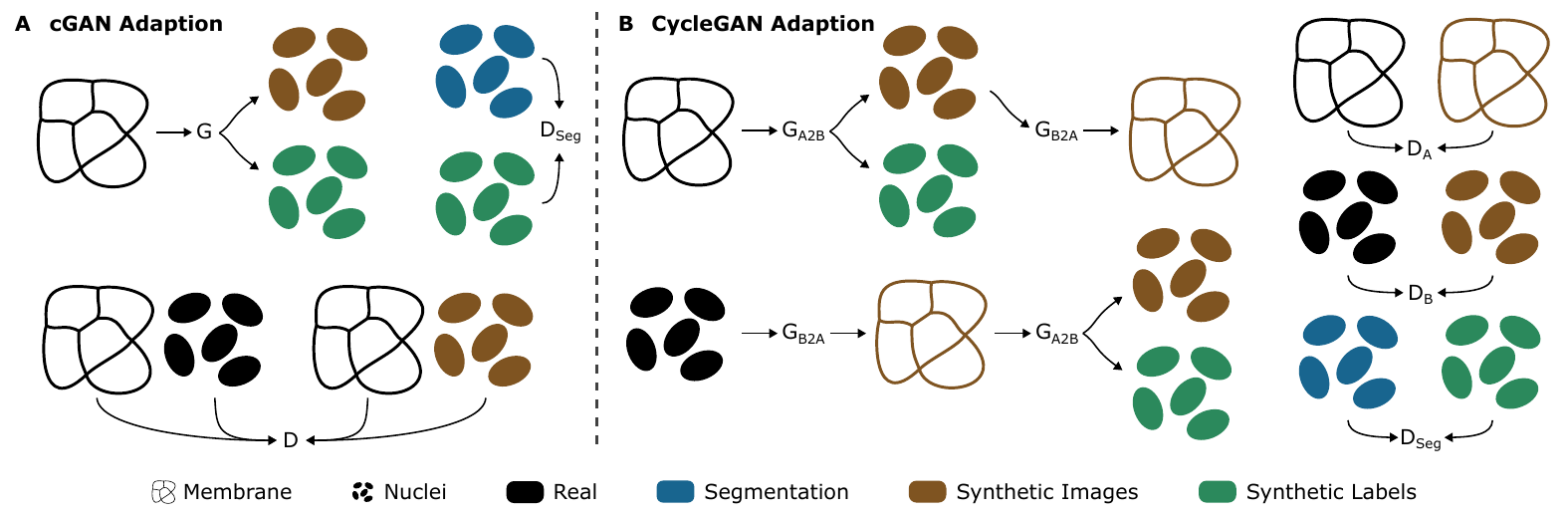}
                \caption{Overview of two proposed GAN training schemes for the generation of not only synthetic images, but also matching labels. Both methods utilize an additional discriminator $D_{Seg}$ to generate additional labels. \textbf{A} and \textbf{B} show the adapted training schemes for a cGAN and CycleGAN, respectively.}
                \label{fig:method_overview_gan_label}
            \end{figure}

\begin{figure}
                \centering
                \includegraphics[width=1\linewidth]{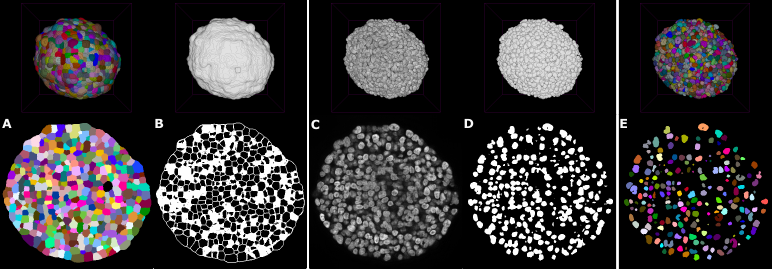}
                \caption{Generation of synthetic images and labels using GAN-based methods. The simulated cell borders (A) are converted into binary membrane signals (B) which are then utilized by the generator to produce synthetic nuclei signals (C) along with corresponding binary labels (D). Subsequently, a post-processing step is employed to extract instance labels (E) from the generated binary labels.}
                \label{fig:method_gan_label_pp}
            \end{figure}

        \paragraph{Adapted GAN training for the additional generation of matching labels}
            Mem2NucGAN-P and Mem2NucGAN-U utilize GANs to obtain nuclei images based on membrane like signals. For the training of segmentation models, corresponding label data to the generated image data is required. However, conventional GAN training methods for image generation lack the ability to output corresponding labels alongside the generated images. To circumvent this constraint, existing approaches resort to generating synthetic labels separately, such as through random ellipse placement, before using GANs to transform them into image data. However, this approach is not feasible for our task, as we aim to transform between two distinct image modalities — membrane to nuclei (\autoref{fig:method_gan_label_pp}b+c) — while only having access to membrane labels via biophysically simulated cell borders (\autoref{fig:method_gan_label_pp}a). Hence, we require the GAN to produce both synthetic nuclei signals and their corresponding labels (\autoref{fig:method_gan_label_pp}c+d). To our knowledge, no existing training scheme facilitates the additional generation of corresponding labels. Therefore, we introduce a novel GAN structure to generate nuclei images along with matching label masks. The new structure can be applied to both cGAN and CycleGAN (see~\autoref{fig:method_overview_gan_label}). First, the adaption of the cGAN is described, while later the methodology is transferred to the CycleGAN.

            The last layer of the generator $G$ is altered to generate a two channel image. The first channel contains the nuclei signal prediction, while the second channel contains a binary label mask corresponding to the nuclei signal in the first channel (see \autoref{fig:method_gan_label_pp}c+d). As the two output channels are based on the same decoder path, i.e., sharing the same feature maps, the alignment between the image and label data is promoted. During training, the discriminator $D$ receives fake samples consisting of the condition (i.e., the membrane signals) and the first channel of the generator output (i.e., nuclei signals) and real samples consisting of the condition (i.e., the membrane signals) and real nuclei signals.
            
            In addition to the traditional training scheme, an extra discriminator $D_{Seg}$ is introduced. While the default discriminator only penalizes the quality of the nuclei channel, the purpose of the second discriminator is to penalize the generated binary label mask. The discriminator $D_{Seg}$ itself is trained to distinguish between real and fake binary labels. We want to emphasize that, in contrast to the regular discriminator $D$, $D_{Seg}$ receives no additional condition. This enables the use of arbitrary, non-paired binary label masks for the training of $D_{Seg}$, diminishing the demand for high-quality label masks in two key aspects. Firstly, with the absence of a corresponding nuclei signal, any overlooked nuclei within the label masks become less conspicuous, thereby minimizing confusion during model training. Secondly, the use of a binary label mask eliminates issues regarding under- and over-segmentation errors.

            As the generator $G$ produces binary labels masks (\autoref{fig:method_gan_label_pp}d), a post-processing step is required to derive instance labels (\autoref{fig:method_gan_label_pp}e) required for segmentation training. In a first step, the predicted label mask undergoes a thresholding procedure to obtain a true binary label mask. Afterwards, the binary membrane signal which is used as input for the generator is subtracted from the binary label mask to split the labels. Then, a binary opening operation is performed to remove potential artifacts. Utilizing the simulated cell label mask, each foreground voxel in the binary label mask is assigned the corresponding cell label. Thereby, foreground voxels in the binary label mask that are zero in the cell label mask, i.e., artifacts, are automatically removed. 
            
            The CycleGAN structure is altered similarly to the cGAN. The generator $G_{A2B}$ is tasked with the generation of a two channel output based on a single channel input featuring a binary membrane. The first channel represents the nuclei signal, while the second channel represents the binary labels. Since there are no segmentations available for real nuclei images, the generator $G_{B2A}$ only receives a nuclei image and no label mask. Similarly, the discriminator $D_B$, only assesses nuclei images to distinguish between real and fake samples. To penalize the generated label masks, an additional discriminator $D_{Seg}$ is introduced, evaluating real and fake binary label masks. As only unpaired binary label masks are required, the same benefits as for the adapted cGAN apply here.

    \subsection{Quality evaluation measures}
        
        \subsubsection{Kernel Inception Distance (KID)}
            To assess the similarity of 3D microscopic images generated by the presented methods to their real counterparts, we employ the Kernel Inception Distance (KID) \cite{Binkowski2018}. As the KID is based on a pretrained Model, designed to process 2D images, the measure cannot be directly applied to 3D images. We therefore calculate the KID measure slice by slice. The mean results for all three major plane directions ($xy$, $xz$ and $yz$) are used to calculate a final mean value of the KID measure.
            
        \subsubsection{Segmentation}
            Raw confocal data sets were converted to multichannel TIFF files using Fiji [36]. 3D segmentation of nuclei and membrane staining (SIR-actin) was performed using Cellpose (V2.2), a deep learning-based instance segmentation tool [37]. To prepare the hand-annotated training data, spheroids of each cell type were first pre-segmented by the pre-trained \emph{nuclei} and \emph{cyto2} models in Cellpose, using the 2 fluorescent markers for DAPI and SiR-actin, as an initial step. From these, 3 patches for each marker with a size of \qtyproduct[product-units = power]{32 x 128 x 128}{\px} (z, y, x) were extracted and manually corrected using the Segmentor software [38]. The location of the patches were selected to feature different image qualities. The patches represent the top, middle, and bottom region of the spheroid, where each patch shows a reduced signal-to-noise ratio. Supervised training from scratch was performed as described in [39] using the command line interface for Cellpose. Please note, that the sample of the spheroid, used to generate the ground truth, was from a previous experiment. Thus, clearing effectiveness slightly differs between the image samples used to generate synthetic data and the image sample from which the ground truth was generated.

            The suitability of different types of synthetic data for the use as training data for segmentation models is tested by training multiple segmentation models. For each presented approach of generating synthetic data, a 3D StarDist~\cite{Weigert2020} segmentation model is trained with four synthetic images.
            To assess the segmentation performance of the models, we utilize two metrics: SEG (Segmentation) and DET (Detection), as used in the Cell Tracking Challenge~\cite{Ulman2017}. The SEG measure evaluates the accuracy of segmentation by quantifying the agreement between the predicted and ground truth segmentation masks. In contrast, the DET metric focuses on the accuracy of detection, measuring how accurately the model identifies objects of interest within the image, while considering factors of false-positives, false-negatives detections and under-segmented objects. Both measures can range from zero to one, representing worst to best result.
            The performance was evaluated by segmenting the whole image and subsequently cropping the segmented image to match the ground truth patches. A direct segmentation of image crops would lead to many incomplete nuclei, potentially decreasing the segmentation quality. By segmenting the entire image, all nuclei remain fully visible, ensuring a more comprehensive assessment of segmentation performance.
            
        \subsubsection{Statistical evaluation}
            We conducted a one-sided Welch-test to compare differences in mean performance among the tested segmentation models. Each model is evaluated on three distinct ground truth patches ($n=3$). Significance levels were pre-defined: *~$p < 0.05$, **~$p < 0.01$, and ***~$p < 0.001$.

    \subsection{Dataset}\label{Sec:Dataset}
        \subsubsection{Sample preparation}
            HT-29 colon cancer cells (ATCC) were cultured in McCoy’s 5A medium (Capricorn) supplemented with 10 \% FBS and 1 \% Pen/Strep and maintained in a humidified incubator at 37 °C with 5 \% CO2 fumigation. For spheroid generation, cells were detached using Trypsin/EDTA and seeded onto the ULA plates at a concentration of 5 x 102 cells per well.  
            
            Spheroids (n=$10$) were transferred after 4 days culturing to Eppendorf tubes, washed once with phosphate-buffered saline (PBS, Sigma Aldrich), and fixed with $4\%$ paraformaldehyde (PFA, Carl Roth) for $1$ h at $37$°C, followed by two washes with PBS containing $1\%$ FBS for $5$ min each. To remove traces of fixative, spheroids were quenched with $0.5$ M glycine (Carl Roth) in PBS for $1$ h at $37$°C with gentle shaking. Spheroids were then incubated for $30$ min in a penetration buffer containing $0.2\%$ Triton X-100, $0.3$ M glycine, and $20\%$ DMSO (all Carl Roth) in PBS to enhance the penetration of antibodies and nuclear stains. Spheroids were then incubated in a blocking buffer ($0.2 \%$ Triton X-100, $1 \%$ BSA, $10 \%$ DMSO in PBS) for $2$ h at $37$ °C with gentle shaking. Samples were then stained with SiR-actin (Spirochrome, $1:1000$) and DAPI (SigmaAldrich, $1:1000$) ON at $37$ °C in antibody buffer ($0.2 \%$ Tween 20, $10$ µg/ml heparin (both Sigma-Aldrich), $1 \%$ BSA, $5 \%$ DMSO in PBS) with gentle shaking. Samples were then washed $5$ times for $10$ minutes each in wash buffer ($0.2 \%$ Tween-20, $10$ µg/mL heparin, $1 \%$ BSA) with gentle shaking and cleared with FUnGI clearing solution ($50 \%$ glycerol (vol/vol), $2.5$ M fructose, $2.5$ M urea, $10.6$ mM Tris Base, $1$ mM EDTA) ON as previously described [35]. Cleared samples were transferred to $18$ well µ-slides (Ibidi) in the same solution and kept in the microscope room for several hours to allow for temperature adjustment.

        \subsubsection{Imaging and data description}
            Spheroids were imaged using an inverted Leica TCS SP8 confocal microscope (Leica Microsystems CMS, Mannheim, Germany) equipped with an HC PL APO 20×/0.75 IMM CORR objective, $488$ nm, $561$ nm and $633$ nm lasers and Leica Application Suite X software. All image stacks were acquired with comparable settings, using Immersion Type F (Leica Microsystems, RI $1.52$) as immersion fluid, with a resolution of \qtyproduct[product-units = power]{1024 x 1024}{\px}, a z-step size of $1$ µm, a laser intensity of $1 \to 1.5 \%$ and a gain setting of $600$ to avoid overexposure of pixels. All image stacks were acquired with z-compensation to compensate for depth-dependent signal loss.

\backmatter





\bmhead{Acknowledgements}

This work was funded by the German Federal Ministry of Education and Research (BMBF) grant 01IS21062B. RB and RR were funded by the Carl-Zeiss foundation, project DigiFIT. This work was funded by the German Federal Ministry of Education and Research (BMBF) as part of the Innovation Partnership M2Aind, projects M2OGA (03FH8I02IA) and Drugs4Future (13FH8I05IA) within the framework Starke Fachhochschulen-Impuls für die Region (FH-Impuls). This work is supported by the Helmholtz Association Initiative and Networking Fund on the HAICORE@KIT partition. The funders had no role in study design, data collection and analysis, decision to publish, or preparation of the manuscript.

\subsection*{Competing interests}
The authors declare no competing interests.

\subsection*{Author contribution}
R.B. and M.R. developed and implemented algorithms. R.B. and E.N. invented the approach to generate synthetic data. E.N. implemented and performed biophysical simulations. M.V. developed the biological sample preparation and performed subsequent imaging. R.B. designed and conducted the experiments. R.B. and M.R. were responsible for the manuscript. M.R., R.R. and S.S. conceived and supervised the project. All authors discussed the results and implications, participated in writing, and commented on the manuscript at all stages.

\end{document}